# Bending the Curve: Operational Cyber Epidemiology for Ransomware

Stephen V Flowerday[a,*], Nikolay Lipskiy[b], Steven Furnell[c], Callum E Flowerday[d], John Hale[e]

[a]*School of Computer and Cyber Sciences, Augusta University,*
[b]*US Center for Applied Medical AI (CAMA), Atlanta, GA, US*
[c]*School of Computer Science, University of Nottingham, UK*
[d]*Department of Chemistry and Biochemistry, Brigham Young University, US*
[e]*Tandy School of Computer Science, The University of Tulsa, US*

**Corresponding author .E-mail address: sflowerday@augusta.edu (S.V. Flowerday).*

## ABSTRACT

Ransomware is often treated as a detection problem. Yet the incidents with the greatest operational impact unfold like outbreaks: a single foothold propagates through identities, administrative tools, and shared services while responders make time-critical decisions with limited visibility. This paper proposes an operational cyber epidemiology framework that adapts the Susceptible–Exposed–Infectious–Removed (SEIR) model to ransomware incident management. In the cyber SEIR ontology, Exposed represents latent compromise and staging, often overlapping the operational notion of dwell time, in which a foothold exists but has not yet produced confirmed secondary compromise, whereas Infectious denotes active lateral propagation. Grounded in ISO 5477:2023 public health emergency preparedness and response (PH-EPR) information management guidance and the United Nations Office for Disaster Risk Reduction and International Science Council (UNDRR-ISC) 2025 Hazard Information Profiles, the framework defines interoperable ransomware case definitions and Essential Elements of Information (EEIs) to support cross-incident comparison. We treat the basic and effective reproduction numbers ($R_0$, $R_e$) as directional, near-real-time decision aids for security operations centers (SOCs) and explicitly separate propagation state (SEIR) from observation status (detected/undetected) to avoid conflating detection capability with spread dynamics. Using publicly reported incidents (WannaCry, NotPetya, SolarWinds, and MGM/Caesars) for illustration, we show how outbreak-style measurements translate telemetry into earlier decisions about isolation, credential containment, and restoration sequencing. We also derive simple protection threshold heuristics (targeting $R_e < 1$) and provide a tool agnostic playbook card that links EEIs to explicit action triggers. The framework's primary contribution is a shared operational language that connects technical telemetry to containment decisions under resource constraints.

## 1. INTRODUCTION AND BACKGROUND

Ransomware is among the most disruptive threats facing modern organizations because it undermines the core functions that information systems are built to provide. It can take services offline, corrupt or destroy data, and interrupt critical workflows. In a five-year dataset of publicly reported incidents (2019–2023), ransomware accounted for roughly 32% of reported cyber events and 38% of reported losses, with prevalence rising sharply over the period (Cyentia Institute, 2024). Once an attacker gains a foothold, compromise rarely stays on a single device; it propagates through identity infrastructure, remote administration tools, shared storage, and software distribution channels. The consequences extend beyond direct financial loss to operational paralysis, reputational damage, and, in some settings, elevated safety risk. For security operations and audit teams, the immediate task is to slow propagation, protect evidence and recovery paths, and restore essential services without reintroducing compromise. This urgency is reflected in the World Economic Forum's Global Cybersecurity Outlook 2025 survey, where 45% of respondents ranked ransomware as their most concerning organizational cyber risk (Joshi et al., 2025).

From an information systems perspective, ransomware is a sociotechnical event. Vulnerabilities and malware matter, but so do asset visibility, patch cadence, access governance, staffing, third-party dependencies, and leaders' ability to coordinate a response. When incidents move quickly, teams need a shared language that links observations to decisions. In practice, critical information is dispersed across threat reports, after-action summaries, and tool-specific dashboards, confounding coordinated decision-making.

Two incidents from 2017 illustrate why an outbreak lens is useful. WannaCry erupted on May 12, 2017, and spread rapidly through a wormable vulnerability addressed in Microsoft's MS17-010 bulletin (Cloudflare, n.d.-b; Microsoft, 2017). NotPetya followed on June 27, 2017. Although often grouped with ransomware, NotPetya is widely described as a destructive wiper masquerading as ransomware because it lacked a workable decryption path even for paying victims (Cloudflare, n.d.-a; Greenberg, 2018). It was seeded through a compromised software update and then moved laterally through multiple mechanisms, including credential misuse and remote execution tools (ESET Research, 2017; Greenberg, 2018). Both incidents caused large-scale disruption. NotPetya produced aggregate losses widely estimated at more than $10 billion, including roughly $250–300 million reported by Maersk alone (Greenberg, 2018; Leovy, 2017; Maersk, 2018).

Taken together, these incidents show how a single foothold can produce both rapid, worm-like spread and slower, stealthier propagation through trusted channels. The dynamics are not biological, but the spread logic is comparable: growth, delay, and threshold effects. Epidemiology offers a mature vocabulary for describing propagation, identifying early warning signals, and evaluating interventions (Anderson & May, 1991; Hethcote, 2000).

Public health emergency preparedness and response (PH-EPR) faces similar coordination challenges across diverse hazards, relying on a mature toolbox of analytical and operational methods. This study draws on three foundations: (1) the United Nations Office for Disaster Risk Reduction (UNDRR) Sendai Framework's all-hazards risk-reduction doctrine, (2) the UNDRR and International Science Council (ISC) hazard taxonomy (UNDRR-ISC), and (3) the PH-EPR information management guidance formalized in ISO 5477:2023. ISO 5477 provides PH-EPR information management business rules and a framework for interoperable vocabularies, standardized case definitions, and measure-driven situational awareness to support emergency decision-making. Applied to ransomware, outbreak constructs provide an operational vocabulary for case definitions, measurement, and decision thresholds. They help response teams reason about spread dynamics and capacity-limited containment for a technical hazard (ISO, 2023; UNDRR, 2015; UNDRR-ISC, 2025).

Against this backdrop, the **Research Question** is: *How can epidemiological outbreak metrics, specifically compartmental models and reproduction numbers, optimize decision-making during resource-constrained ransomware response?*

This question is practical: on the first day of a ransomware incident, leaders tend to ask the same questions:

- Is the trajectory worsening or stabilizing?
- What is driving spread right now?
- Which actions, taken now, will most reduce further propagation?
- What should change to prevent recurrence?
- How do we explain trade-offs to nontechnical decision-makers?

Many response guides describe sound practices, but they do not always connect those practices to a simple model of growth, delay, and operational thresholds. Without a shared model, teams can overreact in one area and underreact in another. Outbreak constructs provide a disciplined vocabulary for diagnosing transmission pathways and selecting control levers. Concise pathogen archetypes (e.g., measles-like rapid contact spread versus cholera-like common source seeding) are used only to keep the analogy operational, not to claim biological equivalence (Brhane et al., 2024; Guerra et al., 2017; Masters et al., 2024).

This outbreak management lens, grounded in PH-EPR practice and simple outbreak metrics, is meant to complement, not replace, engineering rigor and threat intelligence. It offers a compact logic for prioritization. When compromise spreads autonomously through a shared vulnerability, patching and exposure reduction typically yield the fastest gains. When spread is credential-enabled, identity containment and controls that limit lateral movement take priority. When response capacity is constrained, the same intrusion can become far more severe because infectious systems remain able to propagate for longer. These principles are familiar in emergency preparedness and epidemiology, and they translate cleanly to operational cyber decisions.

Public health matured through shared measures, shared terminology, information management and operational standards, rehearsed playbooks, and a culture of learning from events. Epidemiology therefore provides tools to quantify spread, anticipate risk, and justify interventions that support coordinated response across organizations.

Ransomware response is improving, but capability remains uneven across organizations and sectors. Many teams still lack reliable asset inventories, tested backups, prepositioned isolation controls, and clearly defined roles for incident command. The gap is not only technological; it reflects uneven operational maturity. An outbreak lens helps because it offers a consistent way to describe growth and a simple operational aim: drive effective spread below a sustainable threshold.

Epidemiology also links response to preparedness through structured capability descriptions. In public health, for example, the CDC's Public Health Emergency Preparedness capabilities describe how to organize surveillance, information management, incident command, and recovery planning (Centers for Disease Control and Prevention, 2018). ISO 5477:2023 provides PH-EPR informatics guidance that underpins this paper's information management approach. It formalizes all-hazards emergency management principles, standardized case definitions, interoperable vocabularies, and measure-driven situational awareness for managing man-made and technical hazards. We reuse that structure as the backbone for operational ransomware modeling, operationalizing concepts such as susceptibility, transmission, and reproduction number within established PH-EPR doctrine rather than relying on metaphor alone. In applying this established PH-EPR architecture to ransomware threat modeling, this paper contributes to the progressive formalization of cyber epidemiology — moving the field from conceptual analogy toward a standards-grounded, operationally deployable science.

## 2. RELATED LITERATURE AND THEORY

This paper integrates three complementary strands into a single operational framework. Epidemic modeling provides simple, interpretable stages and rates for propagation and delay. Cybersecurity research explains the mechanisms and constraints that shape ransomware incidents, including how detection and mitigation practices change spread pathways. PH-EPR preparedness guidelines explain how organizations convert observations into coordinated action through hazard taxonomies, standardized case definitions, Essential Elements of Information, and incident command. Together, these strands motivate an operational analytics layer that translates technical observations into a common operating picture: pathway, timing (the exposed period), and spread sustainability (Re). That operating picture supports phase-appropriate decisions for isolation, credential containment, and safe restoration.

- Epidemic models: Classic compartmental models, beginning with the SIR model introduced by Kermack and McKendrick (1927), represent a population as groups that move between states. The SEIR family adds an exposed state to represent a delay between infection and infectiousness. Other variants capture reinfection or severe outcomes. These models exhibit threshold behavior and support formal definitions of the basic reproduction number, $R_0$, often computed using next-generation methods (Diekmann et al., 1990; van den Driessche & Watmough, 2002). Modern epidemiology also studies spread on networks and metapopulations, where contact structure shapes outbreak curves and where targeted protection can focus on highly connected nodes (Keeling & Eames, 2005; Keeling & Rohani, 2008). Transmission models are widely used to explore how interventions change trajectories (Centers for Disease Control and Prevention, 2025; Hethcote, 2000). These features, namely clear stages, delay, and threshold behavior, are precisely what this paper reuses through a cyber SEIR mapping and an $R_0/R_e$ lens as it treats ransomware as an outbreak process.

- In applying this established PH-EPR architecture to ransomware threat modeling, this paper contributes to the progressive formalization of cyber epidemiology — a field whose foundations trace from early directed-graph epidemic models of computer virus propagation (Kephart & White, 1991, 1993) through network-topology-aware outbreak analysis (Pastor-Satorras & Vespignani, 2001) and genetic epidemiology approaches to host-level cyber risk (Gil et al., 2014), to recent compartmental modeling of self-propagating malware with formally derived reproduction numbers (Chernikova et al., 2023). By anchoring this progression within established public health emergency preparedness standards — ISO 5477:2023 and CDC PHEP capabilities — rather than relying on structural analogy alone, the present paper advances cyber epidemiology from a theoretically grounded but operationally unmoored discipline toward a doctrine-anchored, deployable analytical science.

A central concept is the basic reproduction number, $R_0$: the average number of secondary infections produced by one infectious case in a fully susceptible population. Measles is a canonical high-transmissibility benchmark, with published $R_0$ estimates commonly cited as 12–18 (Guerra et al., 2017). Recent public health work also demonstrates the operational value of updating transmission models in near-real-time during an active outbreak to forecast size and evaluate interventions (Masters et al., 2024). When $R_0$ exceeds 1, outbreaks tend to grow; when it falls below 1, sustained growth becomes difficult. The effective reproduction number, Re, applies the same logic to settings with partial protection and ongoing intervention. The COVID-19 pandemic underscored that modest shifts in spread and delay can produce large differences in outcomes, and that early $R_0$ estimates are uncertain and context dependent (Alimohamadi et al., 2020; Liu et al., 2020; Park et al., 2020).

- Cyber incident analysis: Ransomware research spans technical reporting, lessons learned reviews, and detailed reconstructions. For WannaCry, reporting emphasized the MS17-010 vulnerability, rapid worm-like propagation, and the role of a sinkhole domain in slowing spread (Hutchins, 2017; Microsoft, 2017). Healthcare focused analyses documented operational consequences, including disruption to the National Health Service (UK) and associated costs (Ghafur et al., 2019; Smart, 2018). For NotPetya, reporting emphasized software supply-chain seeding, multiple internal propagation mechanisms, and destructive outcomes that resembled a wiper more than a profit-seeking extortion event (ESET Research, 2017; Greenberg, 2018; Kaspersky Lab, 2017).

- Cyber epidemiology is a growing body of work that models self-propagating malware with epidemic-style dynamics, derives reproduction numbers, and fits parameters to trace data. Chernikova et al. (2023), for example, show that models with an explicit dormant state can fit WannaCry traces better than simpler formulations. Related work introduces quarantine or isolation states and derives $R_0$ using next-generation methods (Awasthi et al., 2023; Zhou et al., 2023). Overall, this literature supports using SEIR-style stages as decision aids rather than as purely abstract compartments. It also motivates extensions that represent dormant positioning or quarantine as distinct states when modeling ransomware operations and containment.

Beyond compartment models, other work adapts epidemiological constructs such as the epidemiology triad (agent, host, environment) to guide malware control selection and control expenditure decisions, emphasizing sociotechnical context (Flowerday et al., 2024).

Much of the ransomware literature advances specific mechanisms. It includes detection approaches that use dynamic analysis and explainability, and mitigation techniques that aim to slow or confound harmful file activity (Berardi et al., 2023; Gulmez et al., 2024). That work is essential and strengthens the security toolkit. The approach in this paper complements that line by focusing on the operational layer between technical controls and incident command. It links early signals to a small set of outbreak metrics and decision thresholds, helping teams choose interventions that reduce spread while protecting evidence and restoration paths.

- Preparedness and resilience frameworks emphasize standardized hazard taxonomies and case definitions so incidents can be compared across time, sectors, and jurisdictions. The UNDRR Sendai Framework supplies the all-hazards disaster risk reduction doctrine that legitimizes treating ransomware as a technical hazard with population-level consequences. The UNDRR-ISC hazard classification provides the taxonomy and codes needed for consistent aggregation and reporting. ISO 5477:2023 provides PH-EPR information management guidance for interoperable vocabularies, standardized case definitions, and measure-driven situational awareness. At the operational level, MITRE ATT&CK complements these frameworks by characterizing adversary behaviors and transmission dynamics, enabling validation of outbreak-like spread patterns. Together, these frameworks position ransomware within established disaster risk and preparedness paradigms (ISO, 2023; UNDRR, 2015; UNDRR-ISC, 2025).

Three themes shape the operational lens. First, the outbreak vocabulary must remain usable for responders: it should support decisions under fatigue and time pressure, treating $R_0$ and $R_e$ as interpretable directional guides rather than precise predictors. Second, the framework must connect mechanisms to measurement by representing exposure and detection delay explicitly, so observable rates and timestamps can trigger isolation, credential containment, and restoration sequencing before growth accelerates. Third, organizational networks are highly heterogeneous, so a small number of hubs, such as identity providers, software deployment systems, and privileged access pathways, can dominate outcomes; protecting these hubs often yields the largest reduction in effective spread (Cross et al., 2007; Keeling & Eames, 2005). Across themes, effective interventions reduce population exposure, raise protection, and shorten the time compromised systems remain able to transmit.

## 3. METHODOLOGY

This paper is anchored in operational practice. It includes an empirical illustration using publicly available sources. We treat major incidents as case narratives and extract a minimal set of observable milestones such as the first known exposure or compromise window, time of recognition, and onset of visible disruption. From these milestones we derive coarse, comparable measures that map to the framework, including the approximate length of the exposed period and the dominant transmission pathway. The goal is not precise measurement; it is to show that the framework can be applied to real incidents while preserving nuance and producing comparable summaries.

Case-based synthesis using public sources is standard in security operations research because detailed telemetry is rarely public and disclosures are selective and often sanitized or redacted. It is appropriate here because the framework depends on coarse, comparable metrics and decision thresholds rather than fine-grained parameter estimation. When sources diverge, we report ranges or qualitative descriptors. Table 1 summarizes the cases to show how the shared ontology and metrics compare transmission pathways and operational consequences across incidents.

The sources for Table 1 draw on official advisories and credible public reporting. For WannaCry and NotPetya, we rely on incident reporting and technical summaries (Cloudflare, n.d.-b; ESET Research, 2017; Greenberg, 2018; Microsoft, 2017). For SolarWinds, we use government and vendor advisories that document the extended compromise window and remediation guidance (Biasini, 2020; Canadian Centre for Cyber Security, 2020; Cybersecurity and Infrastructure Security Agency, 2021; FireEye, 2020). For MGM and Caesars, we use public incident summaries and disclosures that emphasize social engineering and operational impact (Caesars Entertainment, Inc., 2023; Jones, 2023; MGM Resorts International, 2023).

**Table 1: Illustration of outbreak concepts using public incident narratives**

| Case | Dominant pathway | Exposed period (coarse) | Primary operational impact | Framework lesson |
|---|---|---|---|---|
| WannaCry (2017) | autonomous exploit spread | Hours to days | Rapid service disruption, large infection counts | Patch coverage and exposure reduction dominate early containment ($R_0$ high). |
| NotPetya (2017) | supply-chain seeding | Days | Enterprise wide rebuild, destructive behavior | Trust relationships and admin tools can amplify spread, so segments and contain privileged paths quickly. |
| SolarWinds Orion (2020) | supply-chain seeding | Months | Stealthy compromise, long eviction and assurance work | Long exposed periods require strong detection and disciplined eradication before restoration is reconnected to trusted services. |
| MGM and Caesars (2023) | Social engineering and identity pathway | Days to weeks | Operational disruption, data theft risk | Help desks and identity systems behave like hubs, so verification and privileged access containment matter as much as patching. |

Table 2 defines standardized ransomware case types (suspected, probable, confirmed, removed) using epidemiologic classification within a PH-EPR framework formalized by ISO 5477. We operationalize these definitions using ISO 5477 and the UNDRR-ISC Hazard Information Profiles (HIPs) 2025 update (ISO, 2023; UNDRR-ISC, 2025). In the 2025 HIPs, availability, impact hazards such as disruption and outage are consolidated under Denial of Service (TL0104) (PreventionWeb, 2025b), while Data Breach and PII Breach is coded as TL0102 (PreventionWeb, 2025a).

**Table 2: UNDRR-ISC aligned ransomware case definitions***

| UNDRR-ISC Hazard Code(s) | Case Type | Definition | Case Criteria | Epidemiologic Interpretation |
|---|---|---|---|---|
| TL0101 (Malware) | Suspected | An information system or organizational unit exhibiting early indicators consistent with ransomware activity, without confirmed encryption or extortion. | One or more of the following: (a) detection of ransomware-associated malware components or behaviors; (b) anomalous permission changes, encryption tooling, or ransomware deployment scripts staged but not executed; (c) credential-access, privilege escalation, or remote service probing consistent with pre-positioning for ransomware; (d) alerts indicating ransomware precursors without observable operational impact. | Represents a case under investigation that is consistent with exposure and early-stage positioning. Used to characterize the population at risk and trigger early outbreak monitoring. Not treated as a confirmed source of secondary transmission for $R_0/R_e$ estimation until upgraded to probable/confirmed. |
| TL0101 (Malware) + TL0104 (Denial of Service) | Probable | An information system or organizational unit with confirmed ransomware execution and operational disruption, without verified data exfiltration or external extortion confirmation. | All of the following: (a) confirmed encryption of systems or data consistent with ransomware behavior; (b) observable operational disruption or service degradation; (c) evidence of internal propagation or lateral movement; (d) no confirmed external data release or extortion demand. | Represents an infectious case capable of generating secondary cases. Primary contributor to $R_0$ estimation, reflecting transmission potential prior to containment. |
| TL0101 (Malware) + TL0104 (Denial of Service) + TL0102 (Data Breach & PII Breach) | Confirmed | An information system, organizational unit, or sector level entity experiencing encryption based disruption combined with verified extortion activity, with or without data exfiltration. | All of the following: (a) encryption of critical systems or data; (b) verified extortion demand or ransom note; (c) evidence of command and control activity and/or data exfiltration; (d) documented outage, service interruption, or cascading organizational or sectoral impact. | Represents a severe infectious case with high transmission potential and systemic consequences. Used to model super spreading events, cascading failures, and sector-level $R_0$ dynamics. |

| TL0101 (Malware) (+ applicable TL0104/TL0102) | Removed | An information system or organizational unit previously classified as suspected, probable, or confirmed that has been isolated/evicted and restored to a trusted operational state, with no evidence of ongoing ransomware execution or propagation. | All of the following: (a) system isolated and rebuilt/reimaged or restored from known good backups; (b) malware persistence mechanisms removed and verified; (c) affected credentials rotated/validated as appropriate; (d) reconnected only after meeting reentry criteria and showing no recurrence during heightened monitoring. | Represents removed/recovered cases no longer capable of generating secondary cases. Used to track containment and recovery progress and to estimate removal rate during response. |
|---|---|---|---|---|

**The same ransomware event may progress from suspected to probable to confirmed as evidence accumulates and impacts escalate, and then to removed after isolation/eviction and validation. This mirrors infectious disease surveillance, where case status reflects increasing certainty while the underlying epidemiologic state progresses from exposure to infectiousness and, eventually, removal.*

*Hazard codes follow the UNDRR-ISC Hazard Definition & Classification Review: Technical report (2025) and the 2025 Hazard Information Profiles (HIPs) online reference (UNDRR-ISC, 2025). The 2021 codes TL0022 (Disruption) and TL0023 (Outage) have been consolidated under TL0104 (Denial of Service) in the updated taxonomy.*

In practice, these case definitions make escalation and reporting comparable across incidents by linking observable evidence to a consistent set of surveillance categories.

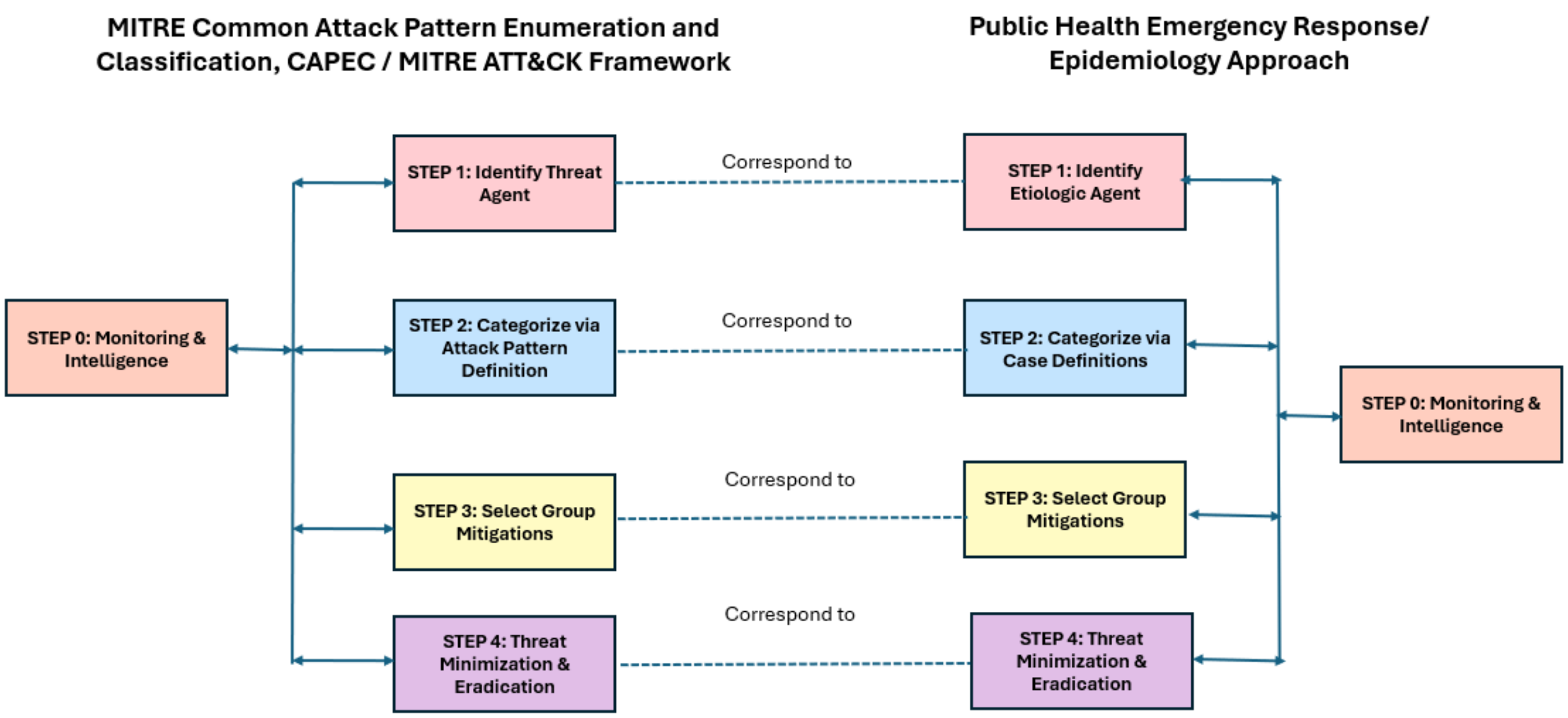

**Figure 1: Comparative case-definition frameworks for biological and cyber threat management**

The main point of Figure 1 is straightforward: standardized classification is not a reporting afterthought; it is what links evidence to phase-appropriate monitoring, mitigation selection, and coordinated containment.

To validate ransomware case classifications, illustrative scenarios were mapped to observable MITRE ATT&CK techniques (Table 3).

**Table 3: Validation of ransomware case definitions using MITRE ATT&CK techniques**

| Case Type | Illustrative scenario | Key ATT&CK Techniques Observed | Validation Logic | Assigned UNDRR-ISC Hazard Code(s) |
|---|---|---|---|---|
| Suspected | Credential theft and tool staging on a single host; no confirmed encryption or extortion. | T1003—OS Credential Dumping; T1078—Valid Accounts; T1059—Command and Scripting Interpreter; T1046—Network Service Scanning (where observed) | Techniques indicate precursor activity and positioning consistent with exposure/staging, without evidence of encryption or extortion to meet probable/confirmed criteria. | TL0101 (Malware) |
| Probable | Confirmed encryption on one or more endpoints with evidence of internal spread; no verified data exfiltration or extortion demand. | T1486—Data Encrypted for Impact; T1021—Remote Services (lateral movement); T1105—Ingress Tool Transfer; T1490—Inhibit System Recovery (where observed) | Presence of T1486 confirms ransomware execution; combined with lateral movement techniques it supports an infectious case designation consistent with probable. | TL0101 (Malware) + TL0104 (Denial of Service) |
| Confirmed | Encryption driven outage across multiple segments plus verified extortion demand and/or evidence of data exfiltration. | T1486—Data Encrypted for Impact; T1490—Inhibit System Recovery; T1041—Exfiltration Over C2 Channel; T1071—Application Layer Protocol (C2 communication) | Encryption plus extortion and/or exfiltration behaviors and cascading service disruption validate a severe infectious case with systemic impact. | TL0101 (Malware) + TL0104 (Denial of Service) + TL0102 (Data Breach & PII Breach) |

Finally, PH-EPR research highlights that outcomes are shaped by capacity as much as by threat characteristics. ISO 5477 formalizes this distinction by separating plans from operational functions, focusing on whether roles are defined, information flows reliably, coordination mechanisms exist, and response and recovery procedures are exercised. Within this framework, epidemiologic modeling serves as a tool for assessing how capacity constraints affect hazard propagation. Consistent with outbreak models, faster removal of infectious nodes reduces overall outbreak size; in cyber terms, rapid isolation and system restoration reduce the period during which system compromise can propagate. With this capacity centered perspective established, we turn to the operational mapping and implications.

## 4. PROPOSED FRAMEWORK

### 4.1 Outbreak framing and SEIR ontology

In an outbreak framing, an organization's enterprise network contains a population of endpoints, servers, identities, and services that can move through stages of compromise:

- Susceptible: reachable and not effectively protected.
- Exposed: foothold established (compromise present) but not yet generating confirmed secondary compromise from this node (no propagation attributable to it). Exposure and infectiousness are propagation states and can exist whether or not the SOC has detected them; detection is tracked as a separate observation status.
- Infectious: able to propagate compromise (e.g., by scanning for vulnerable hosts, abusing stolen credentials, or using remote execution tools), whether detected or not.
- Removed/Recovered: isolated, rebuilt, or otherwise prevented from contributing to further spread.

This explicit mapping keeps the analogy disciplined (see Appendix A for a comprehensive list). Epidemiology distinguishes Susceptible, Exposed, Infectious, and Removed states in the SIR and SEIR family of models (Hethcote, 2000; Keeling & Rohani, 2008; Kermack & McKendrick, 1927). We operationalize the same structure as a cyber SEIR mapping for endpoints, identities, and shared services. Table 4 summarizes the mapping used throughout the paper.

**Table 4: Ontology mapping between epidemiology and ransomware**

| Epidemiology construct | Ransomware or IS construct | Operational implication |
|---|---|---|
| Pathogen or agent | Malware, attacker tools, or intrusion mechanism | Clarifies whether spread is exploit-driven, credential-enabled, or supply-chain-seeded. |
| Host | Endpoint, server, identity, or service | Treat hosts as nodes. Prioritize connected hubs and shared services. |
| Susceptible (S) | Reachable and not effectively protected | Prioritizes patch coverage, exposure reduction, and segmentation to reduce susceptibility. |
| Exposed (E) | Foothold established (compromise present) but not yet generating secondary compromise from this node (no confirmed propagation). | Prioritize shortening time-to-recognition and time-to-containment so exposed nodes do not become infectious. |
| Infectious (I) | Able to propagate compromise, for example via scanning, credential misuse, or remote execution | Focus containment on removing propagation paths and isolating affected segments. |
| Removed or recovered (R) | Isolated, rebuilt, reimaged, or otherwise prevented from spreading | Faster removal reduces total outbreak size and supports restoration safety. |
| Immunity or protection | Patching, hardening, segmentation, least privilege, MFA, and compensating controls | Frames controls as increasing effective protection, which reduces Re. |
| Contact network | Trust relationships, identity links, admin pathways, and management tools | Highlights hubs, privileged paths, and high-connectivity services as priority defenses. |
| Basic reproduction number ($R_0$) | Baseline spread potential in a fully vulnerable segment | Helps reason about which pathways could produce rapid growth if unchecked. |
| Effective reproduction number ($R_e$) | Estimated spread potential after protections and response actions are in place. | Operational target is to drive $R_e$ below 1 so new infections decline. |
| Super-spreaders and hubs | Domain controllers, identity providers, deployment systems, backup orchestration | Supports targeted hardening and tiered administration for high leverage risk reduction. |

| Surveillance | Telemetry, logging, detection engineering, alert triage | Improves observability (detected vs undetected) so early signals trigger action; observability is tracked separately from propagation state. |
|---|---|---|
| Isolation and quarantine | Network segmentation, host isolation, disabling admin tools temporarily | Bends the curve by reducing reachable access and stopping lateral movement. |
| Contact tracing | Path analysis, lateral movement reconstruction, credential impact assessment | Identifies who acquired compromise from whom so containment targets the pathway. |
| Incident command and response capacity | Roles, cadence, communications, restore throughput | Treats capacity as a variable that changes containment speed and severity. |

First, the relevant population is not limited to devices. Identities, administrative tools, and shared services often matter more because they connect many systems; a single privileged account can function like a highly infectious node.

Second, the states are approximations rather than strict categories. A host can be partially exposed, for example when a backdoor exists but has not yet been used for lateral movement.

Third, the ransomware case types described above (suspected, probable, confirmed, removed) support standardized surveillance and reporting, while the Susceptible–Exposed–Infectious–Removed (SEIR) states describe propagation dynamics. Susceptible systems define the population at risk; exposed systems correspond to latent or early suspected cases; and infectious systems align with probable and confirmed cases capable of generating secondary compromise. Using both together helps prioritize prevention, detection, and containment actions.

The exposed period is especially critical because it captures the quiet phase: compromise exists before a clear operational signal forces containment. Responders track two related delays: (1) time-to-propagation (how quickly a foothold produces secondary compromise) and (2) time-to-recognition (dwell time from first compromise to reliable detection and escalation). Adversaries often map the environment, escalate privileges, and stage tools before triggering widespread encryption. Treating this interval as exposed directs investment toward shortening these delays through endpoint and identity monitoring, clear escalation thresholds, and high-signal indicators such as unusual authentication patterns, high-volume remote execution, suspicious use of administrative shares, and abnormal interactions with backup systems.

**4.2 Transmission pathways and case illustrations**

The 2017 ransomware incidents illustrate that propagation follows distinct transmission pathways, each with different implications for outbreak dynamics and response leverage. WannaCry illustrates autonomous (wormable) spread, in which an exploit allowed the malware to propagate without direct human action; early growth was therefore shaped mainly by reachability and patch coverage. Microsoft's MS17-010 bulletin and patching guidance show the intended defensive control, even though many environments remained exposed (Microsoft, 2017). In contrast, NotPetya illustrates a supply-chain-seeded pathway with internal propagation: it originated from a compromised software update and then spread inside organizations through multiple mechanisms, including credential misuse and remote execution tools (ESET Research, 2017; Greenberg, 2018). The practical lesson is to identify the dominant pathway early, because the most impactful controls differ by pathway.

WannaCry also shows why the same malware can produce markedly different outcomes. Networks with strong segmentation, limited legacy protocol exposure, and rapid patch deployment behaved like partially protected populations. Flat networks with delayed patching behaved like largely susceptible populations. An outbreak lens helps explain why some organizations saw a sharp spike while others experienced a contained cluster.

NotPetya further highlights how trusted distribution channels and administrative relationships can act as force multipliers, amplifying systemic organizational vulnerabilities. A compromised update can create many initial exposures, and internal propagation can then exploit the mechanisms defenders rely on for administration, such as remote execution, shared credentials, and centralized management. Incident command may therefore need to make uncomfortable choices, including pausing software distribution, restricting management tools, and temporarily breaking some trust relationships to protect the broader environment (ESET Research, 2017; Greenberg, 2018). The SolarWinds Orion compromise illustrates the same amplified exposure pattern on a larger scale. A trusted update channel can seed many organizations with a backdoor long before any visible disruption, creating a long and difficult eviction problem (Biasini, 2020; Canadian Centre for Cyber Security, 2020; Cybersecurity and Infrastructure Security Agency, 2021).

These recent incidents show that ransomware outbreaks can be driven by identity and process-mediated spread as much as by wormable exploits, with help desks, identity providers, and remote administration platforms acting as effective super-spreaders (Caesars Entertainment, Inc., 2023; Jones, 2023; MGM Resorts International, 2023). Taken together with the 2017 campaigns and SolarWinds, these cases motivate a pathway-based view of ransomware transmission, summarized in Table 5, which maps dominant transmission pathways to their primary $R_0$ drivers and the corresponding control levers most likely to reduce outbreak size.

**Table 5: Transmission Pathways, $R_0$ Drivers, and Control Levers**

| Transmission Pathway | Dominant $R_0$ Drivers | Primary Control Levers |
|---|---|---|
| autonomous exploit spread | Reachability of vulnerable services; patch latency; legacy protocol exposure; short generation time | Rapid patch deployment; exposure reduction; service disablement; network segmentation |
| supply-chain seeding | Size of initially exposed population; trusted update channels; delayed detection; administrative tool abuse | Pause software distribution; revoke trust relationships; credential resets; internal isolation and rebuild |
| credential-enabled lateral movement | Network topology; degree of segmentation; concentration of high-connectivity nodes | Targeted segmentation; protection of high-connectivity segments; prioritized containment |

| Social engineering and identity pathway | Credential reuse; privileged role connectivity; help desk and vendor trust relationships | Strong identity verification; privileged access management; help desk hardening; process controls |
|---|---|---|

### 4.3 Measurement and operational interpretation

Many organizations do not discover ransomware until it has already spread, which is why the exposed state is operationally useful. During the exposed period, compromise is present but response has not yet begun, and lateral movement can resemble routine administration. Shortening this period is high leverage. It depends on endpoint and identity monitoring, careful triage, and rehearsed escalation paths so early signals trigger isolation. SolarWinds illustrates the risk: trojanized updates were distributed months before discovery, leaving many environments exposed long before the event was recognized (Biasini, 2020; Cybersecurity and Infrastructure Security Agency, 2021).

Because exposed periods are often long and invisible, especially in supply-chain and identity-driven incidents, outbreak thinking is operational only when it is anchored in measurement. Within a PH-EPR framework, responders can track a minimal measurement set that approximates spread, recovery capacity, and operational severity: new compromised hosts per hour; newly affected segments or subnets; credential dumping indicators; backup-restore throughput; and service downtime for critical workflows. These measures support thresholds for escalation, including when to disconnect segments, rotate privileged credentials, pause software distribution, or activate continuity plans.

This measurement approach aligns with the Essential Elements of Information (EEI) framework defined in ISO 5477:2023 for public health emergencies: critical information necessary for effective planning, response, and coordination (International Organization for Standardization [ISO], 2023). EEIs are specifically tailored pieces of information that incident commanders require to make time-sensitive decisions. Like public health EEIs, ransomware outbreak metrics must be actionable, timely, and directly linked to decision thresholds. Table 6 presents a minimal measurement set that functions as ransomware EEIs, connecting observable indicators to specific containment and recovery decisions.

**Table 6: Essential Elements of Information (EEIs) for ransomware**

| Operational signal (EEI) | What it indicates | Common trigger for action |
|---|---|---|
| **Exposed phase early indicators (E): detect and interrupt staging** | | |
| failed logins per hour (per segment / identity) | Credential probing, password spraying, or authentication abuse during staging | If sustained or spreading across segments, tighten identity controls, increase triage, and pre-stage isolation of likely impacted segments |
| new service account creations and/or privilege escalation events | Preparation for lateral movement and persistence (service creation, privilege acquisition) and hub targeting | If unplanned service accounts or privileged changes occur, flag for immediate review; freeze privilege changes; initiate privileged credential containment where appropriate |
| remote execution bursts (e.g., | Transition from exposure to infectious lateral movement | If remote execution expands to new hosts, treat as infectious spread and isolate affected |
| WMI/WinRM/PsExec/service creation) across multiple hosts | | segments; suspend high-risk admin tooling where feasible |
| anomalous access to backup/restore systems or security tooling tampering | Positioning to impair recovery or evade detection | If backup or EDR controls are targeted, protect clean environments immediately and segment backup infrastructure from uncertain zones |
| unexpected outbound connections from internal hosts (new destinations/large transfers) | Possible exfiltration staging, command-and-control (C2), or harm escalation during the exposed period | If connections to known bad or anomalous external endpoints occur, block and investigate; treat as potential extortion/harm escalation and coordinate legal/privacy response |
| **Infectious/containment indicators (I): track spread sustainability** | | |
| new compromised hosts per hour (or per cadence) | Spread rate | If rising across cadences, isolate segments and escalate incident command |
| newly affected segments or subnets | Propagation breadth | If a new segment appears, pause inter-segment trust and review firewall and remote management paths |
| credential dumping indicators | Identity compromise risk | Rotate privileged credentials and restrict remote administration paths |
| $R_e$ proxy trend ($R_e$_proxy) | Directional signal of whether spread is accelerating | If $R_e$_proxy > 1 for two cadences, treat spread as sustainable and expand isolation and privileged containment |
| **Removed/recovery indicators (R): track stabilization and safe restoration** | | |
| backup-restore throughput | Recovery capacity | If restore queue exceeds capacity, prioritize critical services and protect clean environment |
| service downtime for critical workflows | Severity | Activate continuity procedures; communicate status and timelines |
| reinfection signals in restored systems | Residual propagation or incomplete eviction | If reinfection signals occur, halt restoration, revalidate clean room, and recheck identity and management-plane compromise |

Measurement should not be limited to technical indicators. Response capacity matters. Staffing, tools, and playbook maturity shape how quickly teams can isolate, rebuild, and restore. In outbreak terms, stronger capacity increases the rate at which infectious nodes are removed from the transmission network. This is one reason preparedness investments often outperform improvisation during a crisis.

### 4.4 Metrics and protection thresholds

We use S, E, I, and R, following standard epidemiologic notation, as shorthand for susceptible, exposed, infectious, and removed (or recovered). The reproduction number $R_0$ describes how much a pathway can spread in a fully vulnerable setting. The effective reproduction number $R_e$ describes spread after protections and response actions are in place. In practice, $R_e$ indicates whether the attack is still spreading under current controls. When $R_e$ stays below 1, spread becomes difficult to sustain.

The SEIR family of models is helpful because it keeps attention on delay and coordination. In epidemiology, the exposed period is the time between infection and infectiousness. In ransomware, it is the time between initial compromise and the point when the attacker can spread reliably through the environment or trigger encryption. A long exposed window creates room for early intervention. A short window leaves little margin and makes containment far harder.

To keep the logic usable, we treat $R_e$ as a practical summary of whether spread is still sustainable. If p is the share of the environment that is effectively protected through patching, segmentation, reduced exposure, or strong identity controls, a useful rule of thumb is $R_e \approx R_0 (1 - p)$. Here, p denotes the effectively protected fraction, and (1 - p) approximates the remaining susceptible fraction. The exact value is rarely known during an incident, but the direction is clear: measures that raise p or reduce reachability push $R_e$ down.

Baseline assumption and limitation: the rule of thumb $R_e \approx R_0(1 - p)$ implicitly assumes a homogeneous (uniformly mixed) contact structure within the modeled network segment, as in classic compartmental models. Enterprise networks are heterogeneous and often hierarchical, so a small number of hubs (identity, software deployment, remote management, backup orchestration) can dominate outcomes.

We therefore use the formula as a baseline heuristic for relatively flat segments and as an operational decision aid rather than a precise predictor. In practice, the qualitative hub guidance in Tables 4 and 5 is how the framework operationalizes heterogeneity: prioritize protection and containment of high-connectivity hubs even when the quantitative proxy is computed at a coarse segment level. Future work will extend the framework with weighted contact networks, centrality-aware coverage, and explicit hub versus leaf distinctions.

A simple example is usually sufficient for decision-making. If a pathway has an $R_0$ near 3 in a mostly vulnerable segment and effective protection is p = 0.6, then $R_e$ is about 1.2. The spread may slow, but it can still grow. If protection rises to p = 0.7, then $R_e$ is about 0.9 and growth tends to decline. That is the practical target that connects technical controls to response decisions.

This pattern is consistent with work that fits epidemiological models to malware traces and derives reproduction numbers from observed propagation (Chernikova et al., 2023). For operations teams, the takeaway is practical. If new infections and newly affected segments keep rising, assume $R_e$ is above 1 and prioritize actions that quickly increase effective protection and reduce reachable access. If those signals fall and remain low, focus on protecting clean environments and restoring services safely without reintroducing susceptible connections.

**Table 7: Illustrative scenarios**

| Scenario | Protection p | $R_e$ (if $R_0$ = 3) | Qualitative outcome |
|---|---|---|---|
| Low protection | 0.2 | 2.4 | Rapid growth unless contained quickly |
| Moderate protection | 0.6 | 1.2 | Slower growth, more time to intervene |
| Higher protection | 0.7 | 0.9 | Growth tends to shrink over time |

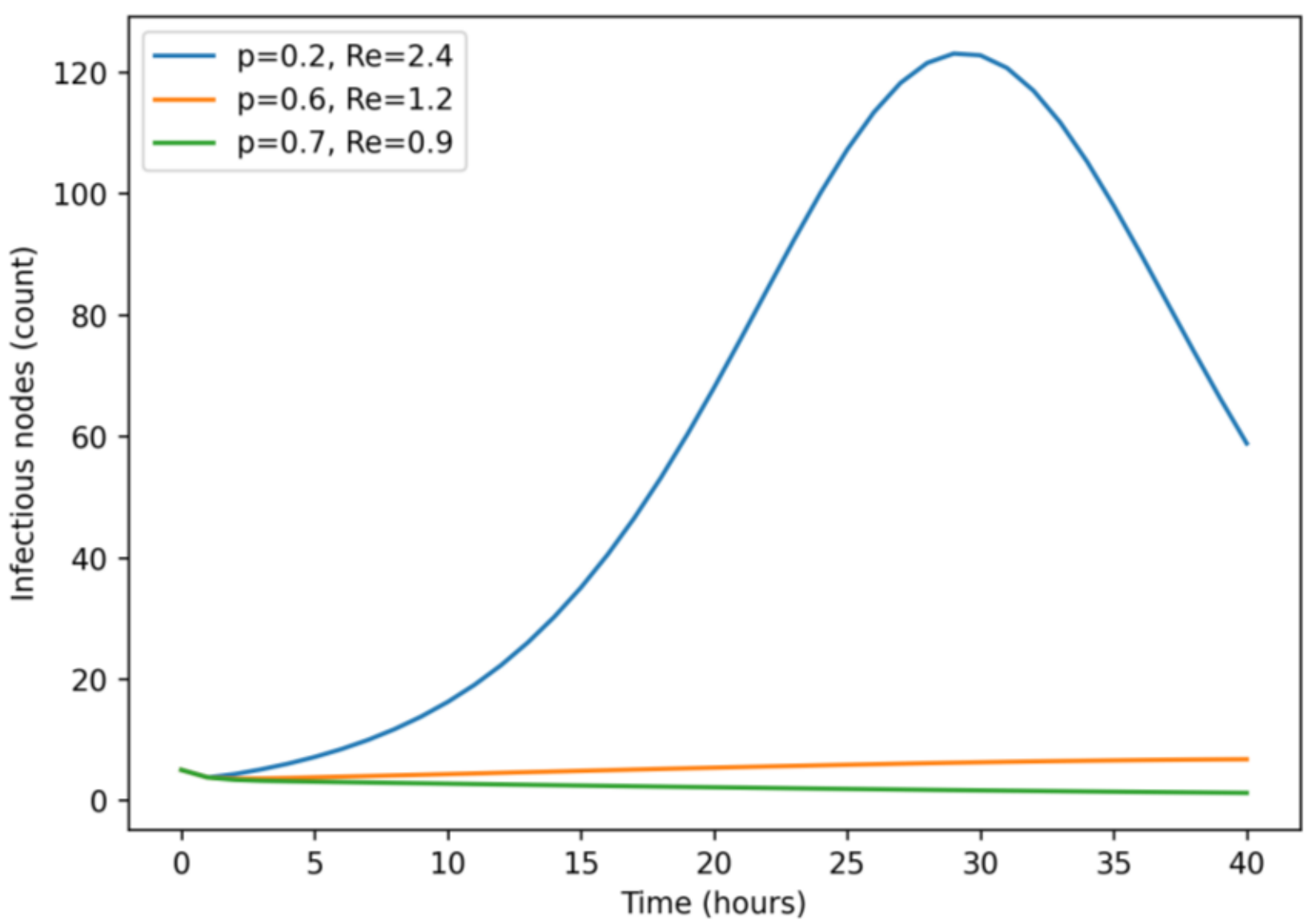

**Figure 2: Illustrative curves of infectious nodes under different protection levels**

### 4.5 Operational use case: estimating spread during an ongoing incident

Retrospective case studies illustrate the framework, but they do not by themselves show how a security operations center (SOC) would use it during an active incident. This subsection therefore outlines an illustrative first day workflow using information typically available in real time: case counts, timestamps, affected segments, and a small set of exposure and lateral-movement indicators.

The goal is not to compute a precise epidemiological $R_0$ in the middle of a crisis. Instead, responders use a directional proxy to judge whether spread remains sustainable (i.e., is $R_e$ above or below 1?) and whether containment actions are bending the curve sooner than a checklist-only approach.

A practical workflow for the first 2–4 hours is:

• Establish a working generation time (g) for the environment (e.g., 30–120 minutes) based on the observed cadence of new hosts showing lateral movement artifacts, ransomware precursors, or confirmed encryptor execution.

• Track an incident time series each cadence: new compromised hosts per g, newly affected subnets/segments per g, and the number of currently infectious nodes (hosts or identities generating lateral movement).

• Compute a sliding-window operational $R_e$ proxy, for example: $R_e$_proxy(t) = new_cases(t) / new_cases(t-g) (optionally smoothed when counts are small). If the rolling $R_e$_proxy stays > 1 for two consecutive cadences, treat spread as sustainable and escalate to broader isolation of inter-segment trust, temporary suspension of remote administration channels, and privileged credential containment.

• If $R_e$_proxy falls below 1 and the number of newly affected segments stabilizes, shift effort from expanding isolation to protecting clean restoration environments and executing staged recovery.

When counts are small (or the prior cadence is zero), responders can apply simple smoothing (e.g., add 1 to numerator and denominator) or treat the signal qualitatively; the operational aim is early directional warning rather than precise parameter estimation.

Illustrative example (conceptual): if the SOC observes 8 new compromised hosts between 09:00–10:00 and 12 new compromised hosts between 10:00–11:00 (with g ≈ 1 hour), then $R_e$_proxy ≈ 12/8 = 1.5. Even with uncertainty in case classification, a sustained $R_e$_proxy > 1 supports earlier network isolation over waiting for widespread encryption because it indicates accelerating propagation.

To avoid tooling lock-in, Table 8 is presented as a functional playbook card (a requirements-style layout) rather than a vendor-specific dashboard mockup.

**Table 8: Conceptual SOC outbreak dashboard/playbook card (based on EEIs)**

| Dashboard / playbook element | What it shows (EEIs) and decision trigger |
|---|---|
| Exposure (E) early indicators | Failed logins per hour (per segment), new privileged group changes, suspicious remote service creation, abnormal admin-tool use. Trigger: two or more indicators rising and unexplained; then increase monitoring, tighten identity controls, and pre-stage segment isolation. |
| Spread & trajectory | New compromised hosts per cadence, newly affected segments, Re_proxy trend. Trigger: Re_proxy > 1 for two cadences or a new segment appears; then isolate inter-segment trust, block lateral-movement paths, and pause high-risk admin tooling. |
| Pathway & hubs | Evidence of dominant pathway (exploit vs. credential vs. supply-chain), plus hub involvement (identity provider, domain controllers, software deployment, backup orchestration). Trigger: hub involvement suspected; then prioritize containment of that hub and rotate or disable associated privileged access. |
| Containment status | Isolation coverage (segments contained/total), privileged credential containment status, high-risk service disablement (e.g., exposed remote services). Trigger: containment gaps in critical paths; then expand isolation and enforce tiered administration. |
| Recovery throughput | Restore queue length, restore rate, clean-environment readiness, integrity checks. Trigger: restore demand outpaces throughput; then prioritize critical services, protect the clean room, and adjust recovery sequencing. |
| Impact & harm (incl. extortion) | Critical workflow downtime, confirmed data staging/exfiltration indicators, regulatory exposure. Trigger: evidence of exfiltration or prolonged downtime; then activate legal and communications workflows and continuity plans in parallel with containment. |

## 5. DISCUSSION AND IMPLICATIONS

### 5.1 Sector context, network effects, and practical impact

Epidemiology considers transmissibility and severity jointly when assessing outbreak dynamics. By analogy, in ransomware, severity can be approximated by operational downtime, data loss risk, recovery time, and secondary harm. WannaCry's impact on the United Kingdom National Health Service illustrates severity tightly linked to service disruption. Retrospective analysis documented disruption to appointments and clinical operations and estimated substantial direct and indirect costs (Ghafur et al., 2019; Smart, 2018). NotPetya shows a different severity pattern: even when the initial infection vector is narrow, internal spread and destructive behavior can create widespread rebuilding needs and prolonged disruption (ESET Research, 2017; Greenberg, 2018).

To keep severity concrete, two simple rates can help. A data fatality rate is the share of affected systems or datasets that cannot be restored to a trustworthy state within a defined window, such as 30 days. A functional impact rate is the share of critical workflows that lose service beyond an acceptable downtime threshold, such as a day for administrative functions or hours for safety-critical functions. These simple measures shift attention from ransom amounts to recoverability and service continuity, which are often what stakeholders care about most.

Outbreak models often assume uniform mixing, but real networks are uneven. Some nodes are highly connected and can function as super-spreaders. In organizational information systems, identity providers, domain controllers, software deployment systems, remote management platforms, shared file services, and administrative hosts often occupy these hub positions. Protecting such hubs is a force multiplier because it reduces both exposure and transmission potential. In practice, this points to strong identity controls, least-privilege administration, separation of duties, and segmentation that limits lateral movement, as recent hospitality sector incidents illustrate (Caesars Entertainment, Inc., 2023; Jones, 2023; MGM Resorts International, 2023). Network-based outbreak models emphasize the same hub effects (Keeling & Eames, 2005).

Targeted protection can be framed as prioritized immunization. If a team cannot patch or harden everything quickly, it can start with the systems that, if compromised, would connect an attacker to the most other systems. In many environments, that includes identity infrastructure, remote management systems, virtualization management, and backup orchestration. Hardening these systems is often decisive because it reduces both rapid spread and the likelihood of catastrophic recovery failure, even when it does not prevent initial compromise.

Limiting credential reuse and enforcing separation between administrative tiers can reduce how far an attacker can travel after a foothold. Similarly, segmentation that keeps critical services and backup infrastructure separated can prevent a local outbreak from becoming an environment-wide event. *These controls do not eliminate risk. They change the shape of the curve.*

Healthcare highlights why ransomware should be treated as more than a financial problem. Disruption can delay care and increase risk. Recent sector guidance increasingly frames ransomware in healthcare as a public health crisis because it can disrupt core clinical and administrative functions and affect patient outcomes (Halcyon, 2025). The National Health Service (UK) lessons learned review emphasizes that cyber incidents can degrade service delivery and require rapid coordination across clinical and technical teams (Smart, 2018). In this context, outbreak-style metrics are useful because they encourage early containment and support clear communication to leadership. The key questions become whether spread is accelerating, stabilizing, or declining and which interventions are most likely to shift that trajectory. Survey evidence from healthcare victims also indicates that recovery can be rapid but costly; in one 2025 survey, 58% of healthcare organizations reported recovering within a week and average recovery costs were about $1.02 million (Sophos, 2025).

Healthcare also illustrates a practical challenge. Technical containment actions can have clinical consequences. Disconnecting a segment might protect the enterprise network, but it can also interrupt workflows. As in public health emergencies, containment actions may shift risk rather than eliminate it, reducing propagation while temporarily increasing operational or clinical risk. This reinforces the value of an incident command structure that includes clinical leadership when necessary. In outbreak terms, some interventions work by changing behavior and workflow, not only technical configuration. Decision quality improves when the team can explain the spread logic and the expected benefit of an intervention in plain language. In safety-critical sectors, ransomware risk is rarely confined to a single organization. Disruption can cascade across interdependent systems (such as emergency services, utilities, transportation, and supply chains), amplifying public health and safety consequences beyond the initial incident.

The same reasoning generalizes to other safety-critical services, including transportation and utilities, where severity includes public impact and safety risk and outbreak-style summaries guide escalation, continuity activation, and trade-offs between short-term disruption and longer-term harm.

**5.2 Applicability to extortion-only and non-propagating events**

Some cyber extortion incidents involve data theft and coercion without self-propagating malware or lateral movement that creates secondary compromised nodes. In these cases, SEIR can still be useful as an operational vocabulary for exposure, response, and recovery, but the reproduction number logic should not be interpreted as a spread metric.

If an event does not generate new compromised nodes through propagation, then $R_0$ and $R_e$ are effectively undefined as outbreak measures because there is no transmission process to sustain. For these non-propagating events, teams should instead use the framework's case definitions, dwell-time metrics, and EEIs to shorten time-to-recognition, confirm containment of access paths, and quantify harm (e.g., data accessed, exfiltration volume, and affected business processes).

For clarity, this paper treats $R_0$/$R_e$ as most applicable when the dominant pathway includes autonomous exploit spread or lateral movement (credential-enabled, tool-enabled, or supply-chain-seeded) that produces secondary compromise. For extortion-only incidents, the same incident command cadence and EEI structure remain useful, but the primary decision thresholds shift from spread control to access eviction, evidence preservation, and impact mitigation.

**5.3 Preparedness, taxonomy, phases, extortion, and incident command**

Public health preparedness frameworks focus on capabilities that translate plans into action. ISO 5477:2023 formalizes this approach for public health emergency preparedness and response (PH-EPR) information systems (ISO, 2023). It defines five core operational functions that apply directly to ransomware response:

- Management: coordination, risk communication, and interagency liaison
- Operations: direct response execution and technical guidance
- Planning: data analysis, forecasting, and resource allocation
- Logistics: resource acquisition, tracking, and service support
- Finance/Administration: cost tracking, budget management, and administrative records

These functions provide a tested organizational structure for outbreak response, whether the threat is biological or digital.

A capability-based approach means organizations can demonstrate operational competencies independent of specific incident scenarios. For ransomware preparedness, this translates to measurable capabilities rather than document compliance:

• Detection capability: Can the organization identify suspicious lateral movement within two hours of initial compromise?

• Containment capability: Can the security team isolate affected segments while maintaining communication channels necessary for incident response coordination?

• Communication capability: Can leadership provide accurate operational status updates to internal stakeholders and external partners hourly during active spread phases?

• Recovery capability: Can the organization validate clean environments and restore systems without reintroducing malware?

• Planning capability: Can the team forecast resource needs, including staffing surge, rebuild capacity, and restore throughput, based on the current spread trajectory?

Each capability maps to the ISO 5477 operational functions. Detection and containment align with Operations. Communication aligns with Management. Resource forecasting aligns with Planning. Recovery validation aligns with both Operations and Logistics. Together, these capabilities transform preparedness from paperwork into operational readiness that can be tested, measured, and improved.

Capability-based preparedness strengthens cross-organizational communication by providing interoperable terms, case definitions, and EEIs that translate technical signals into operational decisions (Centers for Disease Control and Prevention, 2018; ISO, 2023).

A simple test of capability is whether an organization can answer three questions within the first two hours of a suspected incident:

- What is likely affected?
- What pathway is most plausible?

- What immediate containment decision will be taken?

If answering these questions requires a day of debate, the exposed period is effectively extended. When the answers are clear, containment starts sooner and the probability of limiting spread increases. This test reflects a core principle of capability-based preparedness: plans matter less than the ability to execute decisions quickly under uncertainty.

Disaster research reminds us that classification is not bureaucracy; it is how a field accumulates lessons across cases. The IRDR Peril Classification and the EM-DAT taxonomy show how structured vocabularies support comparison and learning (Below et al., 2009; Integrated Research on Disaster Risk, 2014). A practical ransomware taxonomy can follow the same approach using three operational dimensions. First is transmission pathway (autonomous exploit spread, supply-chain seeding, credential-enabled lateral movement, and social engineering and identity pathways). Second is impact severity, ranging from limited disruption with feasible recovery to prolonged disruption requiring large-scale rebuild and, in some cases, destructive outcomes that resemble a wiper. Third is recovery feasibility, ranging from high to constrained to low restoration conditions. Reporting on NotPetya and related variants also highlights how some incidents blend extortion themes with destructive, wiper-like behavior (Symantec Security Response, 2017). Table 9 summarizes the taxonomy and typical control emphases.

**Table 9: Practical taxonomy dimensions for ransomware outbreaks**

| Dimension | Category | Brief description | Example emphasis |
| --- | --- | --- | --- |
| Transmission pathway | Autonomous exploit spread | Wormable exploit, rapid lateral scanning once reachable. | Patch coverage, exposure reduction, segmentation, isolation. |
| | Supply-chain or update seeding | Compromised supplier or update pipeline seeds many targets. | Third-party governance, code signing, distribution, monitoring. |
| | Credential-enabled lateral movement | Stolen credentials plus admin tools, hands-on-keyboard spread. | Identity hardening, tiered admin, least privilege, monitoring. |
| | Exposed remote access or interactive intrusion | Entry via exposed access, then interactive staging and execution. | Surface reduction, MFA, access brokering, logging, path containment. |
| Impact severity | Limited disruption, feasible recovery | Localized impact, bounded downtime and rebuild scope. | Standard playbooks, routine backup restore, focused recovery. |
| | Prolonged disruption, large-scale rebuilds | Enterprise-wide impact, extended downtime, large rebuild effort. | Surge capacity, rebuild orchestration, continuity operations. |
| | Destructive, wiper-like outcomes | Data destruction or recovery path damage, limited restore options. | Immutable backups, crown jewel isolation, protect identity. |
| Recovery feasibility | High | Backups are verifiable, identity trustworthy. | Standard restoration, minimal redesign, reintroduction. |
| | Constrained | Some backups or identity impacted, staged recovery required. | Segmented restore networks, staged credential reset, prioritization. |
| | Low | Backups and management foundations compromised. | Alternate infrastructure, long term continuity, rebuild identity. |

This taxonomy provides a concise way to summarize ransomware outbreaks in terms directly relevant to emergency preparedness, response prioritization, and resilience planning.

Ransomware incidents follow an outbreak-like trajectory: a hidden growth period, an accelerating disruption phase once encryption begins, and a recovery tail shaped by restoration capacity. Organizing the response by phases helps teams match interventions and measurements to that trajectory and capture a consistent set of timestamps for later learning.

The phases are not rigid; they align decisions with what is realistically possible at each point in time. Before an incident, the goal is to reduce susceptibility and exposure. In ransomware terms, this includes patching externally reachable services, removing unnecessary remote access, segmenting critical subnets, hardening identity systems, and testing backups. The strength of this phase is that it is quiet. Teams can make changes deliberately, validate them, and measure coverage.

During early detection, the goal is to shrink the exposed period and hasten containment. When a suspicious foothold is identified, the most valuable time is often the first few hours, before the attacker reaches privileged paths or software distribution systems. It is usually better to isolate a segment early and reverse the decision later than to wait until encryption begins across multiple domains.

During containment, the goal is to reduce effective spread while protecting clean environments. Containment is where $R_e$ is most useful as a decision aid. If new infections and newly affected segments continue, spread remains sustainable. If those signals drop sharply and stay low, containment is taking hold. Containment also requires care so that actions do not destroy the evidence and telemetry needed to understand how spread occurred.

During restoration, the goal is to rebuild safely and avoid reinfection. Outbreak thinking helps because it encourages teams to treat restored systems as susceptible until they are proven clean. Separating restoration networks, validating credentials, and controlling reconnection to uncertain segments reduces the risk of reinfection. The same timestamps used to sketch an epidemic curve can become an improvement plan for patching, segmentation, and escalation speed. In addition to isolation and clean restores, defensive tools can also slow or confound file targeting behavior, buying time for containment and reducing irreversible loss (Berardi et al., 2023).

Modern ransomware incidents combine encryption with data theft and pressure tactics. This changes the severity logic: even if encryption is contained quickly, exposure of sensitive data can extend the incident through legal, regulatory, and reputational consequences. In outbreak terms, an organization may suppress spread yet still face a high-severity outcome. Survey findings in healthcare, for example, show that a meaningful fraction of encryption incidents also involve data exfiltration (Sophos, 2025). This further increases the value of early detection and rapid containment, including behavior-based detectors that can raise signals before the most harmful stages complete (Gulmez et al., 2024).

The outbreak lens remains useful because it separates two questions:

- Is compromise still spreading?
- What harm has already occurred?

Teams should track both. Spread measures guide isolation and credential controls. Severity measures guide communication, legal response, and recovery prioritization.

In practice, this suggests a simple addition to incident command reporting. Alongside the spread curve, teams maintain an exposure and harm assessment that identifies which systems hold sensitive data, whether there is evidence of large-scale exfiltration, and whether backup integrity and restore paths remain trustworthy. This keeps decision-making grounded and reduces the risk that leaders treat a decline in new infections as the end of the incident.

A minimal checklist captures the core levers. Patch externally reachable services and reduce unnecessary exposure. Enforce multifactor authentication and reduce credential reuse. Segment high-value systems and tier administrative privileges. Preposition isolation controls so segments can be contained quickly. Maintain immutable backups with tested restores. Rehearse incident command with thresholds tied to infections per hour and affected segments.

An operational lens must fit inside incident command. A simple approach is to assign responsibility for tracking spread indicators, severity indicators, and response capacity. Those roles inform a small set of decisions, including when to isolate segments, when to rotate privileged credentials, when to pause software distribution, and when to activate continuity procedures. The model functions as a decision aid, not a verdict. Its main benefit is that it keeps the team focused on measurable signals and interventions that directly reduce effective spread.

A practical cadence is to run a short, recurring operations update during the first day. The update can be structured around three questions:

- Is spread accelerating, stabilizing, or declining?
- Which mechanism is most responsible for new infections right now?
- Which next intervention is most likely to reduce spread while protecting recovery paths?

This cadence keeps the team aligned and reduces the temptation to chase every alert. Roles can be assigned in a lightweight way. One person maintains the spread picture, tracking infections, affected segments, and evidence of lateral movement. A second person maintains the recovery picture, tracking backups, clean environments, and restoration throughput. A third person maintains the risk picture, tracking whether identity systems, software distribution channels, or vendor connections are acting as hubs. These roles do not require a large team. They require clarity about what information matters.

Communication becomes easier when the team uses a shared outbreak story. Leaders can understand statements such as “we are seeing expansion into new segments and spread is still sustainable.” They can also understand a near-term containment goal, such as reducing new infections by half over the next two hours through isolation and credential controls. This improves decision quality because proposed actions are tied to a clear logic and an observable outcome.

## 6. CONTRIBUTION

Operational epidemic modeling can serve as an incident management analytics layer when ransomware behaves like an outbreak across identities, administrative tools, and shared services. This paper develops a cyber SEIR ontology that foregrounds early stage delay by distinguishing time-to-propagation from time-to-recognition (dwell time). It operationalizes two interpretable outbreak metrics ($R_0$ and $R_e$), together with a simple protection coverage heuristic, to guide prioritization toward keeping $R_e$ below 1. It also specifies a minimal, auditable measurement set framed as ransomware Essential Elements of Information (EEIs), including exposed phase early indicators (e.g., authentication anomalies and privilege changes) alongside spread, recovery, and severity signals. These EEIs link telemetry to explicit triggers for isolation, credential containment, and restoration sequencing.

Grounded in PH-EPR information management doctrine, the paper operationalizes ISO 5477:2023 case definition and EEI concepts for ransomware using UNDRR-ISC hazard codes. Standard surveillance categories (suspected, probable, confirmed, removed) are defined so they interlock with SEIR propagation states (Susceptible, Exposed, Infectious, Removed), supporting consistent reporting and comparison across incidents without flattening mechanisms. Under this lens, technically diverse events like WannaCry-style worm outbreaks, supply-chain seeding such as SolarWinds, and identity- and process-mediated campaigns such as MGM/Caesars can be summarized using the same ontology, measurement set, and taxonomy. NotPetya is treated as a wiper-like case, reflecting widely reported destructive behavior while preserving its pathway and response lessons.

Alongside the modeling constructs and preparedness frameworks summarized above, the paper also offers a secondary contribution: a communication pattern. Technical observations are condensed into a small set of outbreak-style statements covering the dominant pathway of spread, the estimated exposed period, whether $R_e$ appears above or below 1, whether high-connectivity hubs are implicated, and whether restoration throughput is keeping pace with recovery demand. Together, these statements provide a common operating picture for incident command. They support faster, more transparent decisions under time pressure and help explain trade-offs to nontechnical leaders. The same structure supports disciplined after-action review by pairing the statements with comparable timestamps (first compromise, first detection, first isolation, credential containment, restore start, and critical service restoration). It also supports comparison by pairing them with summary dimensions from the taxonomy (pathway, detection delay, containment time, peak spread rate, severity category, recovery feasibility, and time to critical service restoration). That consistency enables organizations and sectors to benchmark preparedness, accumulate comparable lessons, and improve over time instead of treating each large ransomware incident as an isolated crisis.

## 7. LIMITATIONS AND FUTURE WORK

This paper deliberately simplifies a complex reality. Outbreak data in cyber settings are often incomplete: telemetry is biased toward monitored endpoints, detection delays censor early infections, and disclosure varies across sectors. The contact network is multilayered and changes during response as teams segment networks and disable services. The $R_e \approx R_0(1 - p)$ rule of thumb should therefore be interpreted as a homogeneous-mixing baseline within a segment and used as a directional aid rather than a predictive estimate. In hierarchical or scale-free enterprise networks, a small number of hubs can dominate spread, and centrality-aware protection and weighted-contact models are needed for precise estimation. Attackers also adapt, so key parameters such as effective transmission rate are not stationary. Many outcomes depend on human decision-making, including patch timing, credential hygiene, restoration choices, and delays in establishing incident command. The work focuses on organizational environments. While the same outbreak logic can describe internet-scale worm spread, intervention levers and governance differ materially at the global internet level.

Future work connects mechanistic models to operational data and response decisions. Priorities include (1) standardized incident timelines and minimal data schemas that support cross-case estimation; (2) hybrid models that combine compartment dynamics with explicit network structure (e.g., weighted contact graphs and centrality-aware 'hub versus leaf' coverage); and (3) explicit modeling of detection and reporting delays. A second direction is to couple spread models with response capacity, including staffing and tools, to quantify the operational value of preparedness investments. A third direction is adversary-aware extensions that treat attacker choices as part of the system rather than as exogenous noise. These steps make outbreak analogies more operational by increasing predictive utility while keeping interpretation accessible. SEIRS variants can represent loss of protection or reinfection, and SIDR-style variants can represent dormant or stealth positioning before visible encryption (Chernikova et al., 2023).

## 8. CONCLUSION

Ransomware incidents often behave like pathogen outbreaks. They spread through connected systems, accelerate when exposure is high, and slow when effective protections are in place. The $R_0/R_e$ lens makes this easy to summarize. This is the domain of cyber epidemiology - the formal application of epidemiological theory and public health preparedness doctrine to cyber threat analysis - and it offers a disciplined way to describe outbreak behavior and connect observations to interventions. By keeping the model simple, using $R_0$ and $R_e$ as decision aids, and integrating preparedness capabilities into incident command, organizations can respond earlier and more consistently. The operational goal is to reduce effective spread so growth slows, the environment stabilizes, and critical services can be restored safely.

In answer to the research question posed in Section 1, the paper demonstrates that epidemiological concepts can be adapted to support operational decision-making during ransomware response, particularly under constrained resources (time, staffing, and incomplete visibility) in a way that remains both rigorous and usable.

A cyber SEIR ontology gives teams a shared language for Susceptible, Exposed, Infectious, and Removed states, including the often overlooked dwell-time (staging) window between initial compromise and observable propagation or disruption. Outbreak metrics then turn that language into action: $R_0$ and $R_e$ indicate whether spread is likely to accelerate and whether current protections are adequate. Most importantly, the framework ties familiar security controls such as patching, segmentation, reduced exposure, identity hardening, and rapid isolation to a single operational aim: raise effective protection and shorten the exposed period until $R_e$ stays below 1 and the curve begins to bend. Coupled with a minimal measurement set and incident command roles, the approach supports better preparedness, earlier detection, faster containment, and safer recovery.

In practical terms, the approach supports repeatable learning. If teams capture the same few timestamps and measurements each time, they can see whether detection is getting faster, whether containment starts bending the curve sooner, and whether restoration becomes more reliable. Over time, that evidence supports better investment decisions in patching, segmentation, identity hardening, and recovery engineering. Operational epidemic modeling does not prevent every compromise, but it helps prevent local incidents from becoming enterprise crises. More broadly, each time a team applies this framework and captures its measurements, it contributes evidence to a growing body of cyber epidemiological practice - evidence that, aggregated across organizations and sectors, may eventually support the population-level surveillance, cross-incident benchmarking, and collective defense that mature public health systems have long relied upon.

## APPENDIX A

### Epidemiology & Cyber Epidemiology: Parallel Construct Reference

| Category | Construct | Epidemiology Definition | Cyber Epidemiology Definition | Ransomware / Cyber Example |
|---|---|---|---|---|
| Core Agents & Hosts | Pathogen / Agent | Biological entity (virus, bacterium, parasite, or other pathogen) capable of causing disease in a susceptible host. | Malware, exploit code, ransomware, or other malicious payload capable of compromising vulnerable systems. | WannaCry (S0366), Ryuk (S0446), LockBit 2.0 (S1199) / LockBit 3.0 (S1202) ransomware payloads. |
| Core Agents & Hosts | Reservoir | Natural habitat or population where a pathogen normally lives, multiplies, and persists, serving as a source for transmission to susceptible hosts (e.g., animals, humans, soil, water). | Persistent attacker-controlled infrastructure where malicious code or coordination mechanisms are maintained long-term (e.g., C2 servers, bulletproof hosting, dark-web malware repositories). Note: exploit kits are delivery mechanisms (closer to vectors) rather than true reservoirs. | Ryuk (S0446) / Conti (S0575) C2 infrastructure; dark-web malware repositories; bulletproof hosting servers. |
| Core Agents & Hosts | Vector | A living organism — typically an arthropod (e.g., mosquito, tick, flea) — that transmits a pathogen from one host to another, either biologically (pathogen replicates within the vector) or mechanically (pathogen carried passively on the vector's body or appendages without replication). | Delivery mechanism that passively carries and transmits malicious payloads between systems, analogous to a mechanical vector (no pathogen replication occurs within the delivery vehicle). Examples: phishing emails, malicious downloads, compromised software updates. | Phishing email attachment; malicious link; SUNBURST (S0559) delivered via trojanized SolarWinds Orion update. |
| Core Agents & Hosts | Host | A living organism (human or animal) or population that harbors a pathogen and bears the health burden of infection; individual host characteristics (age, immune status, genetics, nutritional state) modulate susceptibility, disease severity, and outcome. | A single networked system, device, or endpoint — or an organization as a collective — that harbors a threat and bears the direct impact of compromise; host characteristics (OS, patch level, configuration, privileges) modulate exploitability and damage potential. | Hospital server, clinical workstation, EHR database node. |
| Compartmental States | Susceptible (S) | Individuals lacking prior immunity or prophylactic | Systems lacking security protections (patches, authentication controls, | Unpatched Windows systems vulnerable to CVE-2017-0144 |
| | | protection, fully vulnerable to infection upon adequate pathogen exposure. | segmentation) and therefore vulnerable to exploitation upon threat-actor contact. | (EternalBlue, T1210) exploited by WannaCry (S0366). |
| Compartmental States | Exposed (E) | Individuals infected but not yet infectious; in the latent period before pathogen shedding or transmission capacity is established. (Distinct from incubation period, which ends at symptom onset.) | Systems where an initial compromise or foothold has been established but lateral propagation has not yet begun; attacker in pre-lateral-movement dwell phase. | Attacker gains credential access but has not yet launched encryption payload. |
| Compartmental States | Infectious (I) | Individuals actively shedding pathogen and capable of transmitting infection to susceptible contacts, driving population-level epidemic spread. | Systems actively propagating threats to additional systems across networks (e.g., worm propagation, lateral movement via SMB or Active Directory). | Compromised workstation spreading ransomware laterally through SMB or Active Directory. |
| Compartmental States | Removed / Recovered (R) | Individuals who recover with acquired immunity or who die; no longer infectious and no longer contributing to active transmission chains. (Isolation is an intervention acting on the I compartment, not a defining feature of R.) | Systems fully remediated, rebuilt, or decommissioned such that they no longer propagate compromise; no longer participating in active attack chains. | Server restored from clean backup after ransomware removal and reimaging. |
| Protection & Immunity | Immunity / Protection | Natural or vaccine-acquired immunity, or behavioral prophylaxis, preventing infection from establishing despite pathogen exposure. | Security controls preventing successful compromise despite threat exposure (e.g., patching, hardening, endpoint protection, MFA). | Patched systems immune to CVE-2017-0144 (EternalBlue); EDR blocking WannaCry (S0366) / LockBit 3.0 (S1202) execution. |
| Protection & Immunity | Herd Immunity / Network Resilience Threshold | Proportion of immune individuals in a population required to interrupt sustained community transmission ($1 - 1/R_0$); unimmunized individuals are indirectly protected. | Proportion of hardened or segmented systems required to interrupt self-sustaining network compromise spread; unpatched legacy assets indirectly protected through collective defense. | Sufficient patch coverage across enterprise preventing ransomware achieving $R_e > 1$. |
| Transmission Dynamics | Contact Network | Social, spatial, or biological connections enabling pathogen transmission between hosts; network topology (density, | Logical and physical interconnections (shared credentials, trust relationships, API calls, supply chains) enabling threat | Active Directory trust relationships enabling ransomware lateral movement across the enterprise. |
| | | clustering coefficient) shapes outbreak dynamics and determines which individuals are at highest risk. | propagation between systems; topology drives lateral movement paths and blast radius. | |
| Transmission Dynamics | Transmission Route | Mechanism by which pathogen passes between hosts (e.g., respiratory droplet, fecal-oral, vector-borne, bloodborne, contact); route determines exposure risk and informs prevention strategy. | Mechanism by which threats propagate between systems (phishing, drive-by download, lateral movement via SMB/RDP, supply-chain compromise); route determines detection and containment approach. | CVE-2017-0144 / SMBv1 exploit (EternalBlue, T1210) as primary propagation route for WannaCry (S0366). |
| Transmission Dynamics | Incubation / Dwell Period | Time between initial infection and onset of symptoms or signs; longer incubation enables wider silent spread before cases are detected and isolation initiated. | Time between initial compromise and detection (median ~200 days historically); extended dwell enables deeper network penetration, data exfiltration, and persistence establishment before response. | Ransomware operators maintaining network access for weeks before deploying encryption payload. |
| Reproduction Numbers | Basic Reproduction Number ($R_0$) | Average number of secondary infections generated by one infectious case in a fully susceptible population with no interventions in place. | Average number of additional systems compromised by one infected system in a fully vulnerable, fully connected network with no security controls active. | WannaCry (S0366) spreading automatically across unpatched networks via EternalBlue (CVE-2017-0144, T1210); estimated $R_0$ ~2–3 in fully unpatched environments. |
| Reproduction Numbers | Effective Reproduction Number ($R_e$) | Actual average number of secondary infections accounting for pre-existing immunity, behavioral changes, and active interventions; $R_e < 1$ required to terminate epidemic. | Actual secondary compromises accounting for patching coverage, network segmentation, detection speed, and response effectiveness; $R_e < 1$ required to contain breach propagation. | WannaCry (S0366) spread contained after emergency MS17-010 patch reduced $R_e$ below 1; network isolation halted lateral movement. |
| Amplifiers | Super-Spreaders / Network Hubs | Individuals or settings generating disproportionate transmission events due to high contact rates, elevated viral | Critical infrastructure nodes whose compromise amplifies propagation scope (e.g., domain controllers, identity providers, jump | Domain controller compromise enabling Conti (S0575) / LockBit 3.0 (S1202) deployment across entire |

| | | shedding, or environmental amplification (e.g., crowded indoor settings, healthcare facilities). | servers, cloud gateways, CI/CD pipelines). | enterprise in a single operation. |
|---|---|---|---|---|
| Amplifiers | Zoonotic Spillover / Third-Party Risk | Pathogen crossing species barriers from animal reservoir to human population, often initiating a novel epidemic; reservoir control and surveillance at the human-animal interface are primary prevention levers. | Threat entering the enterprise from external supply-chain or vendor vectors, bypassing perimeter controls; vendor security posture and third-party risk management are primary risk levers. | SUNBURST (S0559) delivered via trojanized SolarWinds Orion update, bypassing perimeter controls to reach high-value targets (APT29 supply-chain operation, Campaign C0024). |
| Surveillance & Detection | Surveillance | Systematic, continuous collection, analysis, and interpretation of health data to detect outbreaks, monitor trends, and guide public health action. | Continuous collection and analysis of security telemetry via SIEM, EDR, NDR, and threat hunting to detect indicators of compromise, anomalies, and adversary TTPs. | SOC detecting unusual file encryption behavior via SIEM alerts indicating active Ryuk (S0446) or LockBit 3.0 (S1202) ransomware. |
| Surveillance & Detection | Case Definition | Standardized criteria (clinical, laboratory, epidemiologic) used to consistently identify and classify cases for surveillance and outbreak investigation. | Standardized detection rules and alert thresholds (YARA rules, Sigma rules, behavioral baselines) used to classify true positive security events and reduce false-positive burden. | YARA rule matching LockBit 3.0 (S1202) binary patterns triggering confirmed-case alert in SIEM. |
| Surveillance & Detection | Sensitivity & Specificity | Sensitivity: proportion of true cases correctly detected; specificity: proportion of non-cases correctly excluded. Trade-off shapes surveillance system design. | True-positive rate (detection coverage) vs. false-positive rate (alert noise). Trade-off shapes detection rule tuning — high sensitivity for critical assets, balanced specificity to prevent analyst fatigue. | High-sensitivity ransomware behavioral rules on domain controllers vs. tuned rules on general endpoints. |
| Surveillance & Detection | Outbreak Threshold | Pre-defined case count or rate triggering escalated public health response; defined relative to baseline endemic levels. | Alert threshold or risk score triggering escalated SOC/CSIRT response; defined relative to baseline behavioral norms and asset criticality. | Spike in SMB connection attempts crossing threshold triggering major incident declaration. |
| Containment & Response | Isolation & Quarantine | Separation of confirmed infectious | Network segmentation, endpoint isolation, VLAN | EDR automatically isolating infected |

| | | individuals (isolation) or exposed individuals (quarantine) from the susceptible population to interrupt active transmission chains. | quarantine, or account suspension of confirmed-compromised or suspicious systems to prevent lateral movement and C2 communication. | workstation from network upon ransomware detection. |
|---|---|---|---|---|
| Containment & Response | Contact Tracing | Systematic identification, notification, and monitoring of individuals exposed to a confirmed case to interrupt onward transmission chains and identify secondary cases. | Attack-path reconstruction, forensic timeline analysis, and blast-radius mapping to identify all systems touched by confirmed threat actor and detect secondary compromises. | Tracing which servers communicated with the initially infected endpoint to identify lateral spread. |
| Containment & Response | Decontamination / Remediation | Clinical treatment to eliminate pathogen from confirmed cases, reducing infectious duration and onward transmission potential. | Malware removal, credential rotation, vulnerability patching, and reimaging of compromised systems to eliminate persistence mechanisms and restore clean state. | Full reimaging of encrypted servers and rotation of all compromised credentials post-ransomware. |
| Containment & Response | Incident Command & Response Capacity | Coordinated, multi-agency organizational response to outbreaks structured per ICS/NIMS protocols; spans detection through recovery and after-action review. | Activated CSIRT/SOC response structured per IR playbooks; spans triage through containment, eradication, recovery, regulatory reporting, and post-incident lessons-learned review. | Activation of incident response team, system shutdown, law-enforcement notification, regulatory disclosure. |
| Policy & Intervention | Non-Pharmaceutical Interventions (NPIs) | Behavioral and environmental measures (masking, physical distancing, ventilation improvements) that reduce transmission without pharmacological agents. | Non-technical security controls (security awareness training, phishing simulation, access policies, insider-threat programs) reducing risk without relying solely on endpoint or network tooling. | Mandatory ransomware awareness training reducing successful phishing click rates. |
| Policy & Intervention | Vaccination Campaign / Patch Campaign | Population-level administration of vaccine to induce protective immunity at scale, reducing the susceptible pool below | Enterprise-wide patch deployment, vulnerability remediation, and configuration hardening campaigns reducing exploitable attack | Emergency MS17-010 patch deployment eliminating CVE-2017-0144 (EternalBlue) across enterprise, reducing |
|  |  | the herd immunity threshold. | surface below adversary viability threshold. | WannaCry (S0366)-vulnerable systems below epidemic threshold. |
| Policy & Intervention | One Health / Collective Defense | Integrated approach recognizing human, animal, and environmental health as interdependent; cross-sector collaboration required for effective epidemic prevention and control. | Integrated threat-intelligence sharing across sectors (ISACs, government partnerships, vendor disclosures) recognizing enterprise, industry, and national cybersecurity as interdependent. | FS-ISAC sharing ransomware IOCs across financial sector to enable pre-emptive blocking. |

*Note. Cyber epidemiology examples were verified against the MITRE ATT&CK® knowledge base (MITRE, 2025)*. Software names and identifiers (e.g., WannaCry, S0366; Ryuk, S0446; LockBit 2.0, S1199; LockBit 3.0, S1202; SUNBURST, S0559; Conti, S0575) reflect official ATT&CK software entries. EternalBlue is referenced as CVE-2017-0144 and mapped to ATT&CK technique T1210 (Exploitation of Remote Services), as it does not have a standalone software entry in the ATT&CK catalog. ATT&CK® is a registered trademark of The MITRE Corporation.*

** MITRE. (2025). MITRE ATT&CK® software (v18). The MITRE Corporation. https://attack.mitre.org/software/*